\documentclass[pra,twocolumn]{revtex4}

\usepackage[utf8]{inputenc}

\usepackage{array}
\newcolumntype{P}[1]{>{\centering\arraybackslash}p{#1}}

\usepackage{bm}
\usepackage{amsmath}

\usepackage{amssymb}
\usepackage{graphicx}
\usepackage{multirow}

\begin{document}

\newtheorem{theorem}{Theorem}
\newtheorem{corrolary}{Corollary}

\def\pr{\prime}
\def\be{\begin{equation}}
\def\en#1{\label{#1}\end{equation}}
\def\d{\dagger}
\def\bar#1{\overline #1}
\def\U{\mathcal{U}}
\newcommand{\per}{\mathrm{per}}
\newcommand{\rd}{\mathrm{d}}
\newcommand{\vare}{\varepsilon }

\newcommand{\etab}{\bm{\eta}}
\newcommand{\br}{\mathbf{r}}
\newcommand{\m}{\mathbf{m}}
\newcommand{\s}{\mathbf{s}}
\newcommand{\bk}{\mathbf{k}}
\newcommand{\bl}{\mathbf{l}}

\newcommand{\alphb}{\bm{\alpha}}

\title{Noise in     Boson Sampling   and the threshold of efficient  classical simulatability}

\author{V. S. Shchesnovich }

\affiliation{Centro de Ci\^encias Naturais e Humanas, Universidade Federal do
ABC, Santo Andr\'e,  SP, 09210-170 Brazil }

\begin{abstract}
We study the     quantum  to classical transition in Boson Sampling by  analysing   how $N$-boson interference is affected by inevitable noise in an experimental setup.  We adopt   the  Gaussian noise  model   of Kalai and Kindler for   Boson Sampling and show that  it appears from some  realistic experimental imperfections.       We reveal a  connection   between noise  in Boson Sampling   and    partial  distinguishability of    bosons, which  allows us to prove efficient classical  simulatability of  noisy no-collision Boson Sampling    with   finite noise amplitude $\epsilon$, i.e., $\epsilon = \Omega(1)$ as $N\to \infty$.  On the other hand,  using  an equivalent representation of   network noise   as   losses of bosons  compensated by  random (dark) counts of detectors,  it is proven  that   for noise amplitude inversely proportional to  total number of bosons, i.e., $\epsilon=O(1/N)$,  noisy no-collision Boson Sampling is as hard to   simulate classically as in the noiseless case.  Moreover,  the  ratio of ``noise clicks" (lost bosons compensated by dark counts) to the  total number of bosons $N$ vanishes as $N\to \infty$  for arbitrarily small noise amplitude, i.e., $\epsilon = o(1)$ as $N\to \infty$, hence, we  conjecture  that such a noisy Boson Sampling is  also  hard to simulate classically.    The results  significantly relax  sufficient  condition on noise in a  network  components, such as two-mode beam splitters, for classical hardness of experimental Boson Sampling. 

\end{abstract}
\maketitle

\section{Introduction}

Quantum supremacy \cite{QS}, i.e., the computational advantage over digital computers in some specific tasks,  promised by quantum mechanics to exist already in  mesoscopic-size  devices,       would be a significant step on the way to  the universal quantum computer \cite{BookCC}.   Several proposals were put forward to achieve this goal: Boson Sampling model \cite{AA}, the clean qubit model  \cite{Cqubit} and the commuting quantum circuits model \cite{CSM}. Under  some   computational complexity conjectures,  these models cannot be simulated efficiently  on digital computers (i.e., with computations polynomial  in the model size). The question is whether  the quantum advantage can be maintained under unavoidable imperfections (i.e., noise)  in any experimental realisation  \cite{HQCF}.

In this work we focus on Boson Sampling \cite{AA}, where the classically hard task consists of   sampling from many-body   quantum interference of    $N$ single  bosons on a  unitary linear   $M$-dimensional network,  which  could be  random  \cite{GBS}, but  have to be known  in each run \cite{BSU}.  For  arbitrary $M\ge N$  the quantum amplitudes   of many-body   interference of $N$ bosons are  given   \cite{C,Scheel} by the matrix permanents \cite{Minc} of $N$-dimensional submatrices     of  a unitary  network   matrix  and are generally classically hard to compute     \cite{Valiant,JSV,A1}, with the fastest known algorithm,   due to Ryser \cite{Ryser,Glynn}, requiring   $O(N 2^N)$  operations. For  sampling problem  from the output distribution, in Ref.  \cite{AA}  it is shown that in  no-collision regime,  when the output ports receive at most one boson (i.e., for at least $M\gg N^2$ \cite{Bbirthday}),   classical simulation of the respective output probability distribution for an arbitrary network matrix  to  error  $\varepsilon$ with the computations      polynomial in  $N$ and  $1/\varepsilon$ is  impossible, if  some plausible conjectures are true. 
 In the no-collision regime, $N$-dimensional submatrices  of a Haar-random unitary network matrix are well approximated by  i.i.d. standard complex Gaussians rescaled by $1/\sqrt{M}$  \cite{AA}.  The classical  hardness   of no-collision Boson Sampling is due to the   classical hardness of   approximating, to a multiplicative error,   the absolute value of the  matrix permanent of an $N$-dimensional matrix of  i.i.d. standard complex Gaussians, one of the conjectures of Ref.  \cite{AA}.

In comparison to the universal quantum computation  with linear optics   \cite{KLM} Boson Sampling seems to be    simpler to realise experimentally \cite{E1,E2,E3,E4}, as it does  not require neither  interaction between photons nor error correction schemes.  Some improvements are  available for its  experimental  implementation, such as  Boson Sampling from a Gaussian state  \cite{GBS},  experimentally tested  in Ref. \cite{E5},  and on   time-bin network \cite{TbinBS}, also  tested experimentally  \cite{TbinBSExp}.   Spectacular  advances in experimental Boson Sampling  with linear optics are being continuously reported  \cite{pure1phBS,LossBS,12phBS}. Moreover, alternative  platforms include   ion traps \cite{BSions}, superconducting qubits \cite{BSsuperc,Gst},   neutral atoms in optical lattices \cite{BSoptlatt}  and  dynamic   Casimir effect \cite{BScas}.

Recently  advanced classical algorithms were  shown to be able to  sample from the output distribution of Boson Sampling  on  digital computers up to $N\approx  50$ bosons \cite{QSBS,Cliffords}, a  scale-up of the previous estimate $N\approx 30$ \cite{AA}.  Moreover, unavoidable  imperfections/noise  in an experimental realisation of Boson Sampling can  reduce the gap between classical and quantum simulations, allowing  for  efficient classical algorithms \cite{KK,K1,R1,OB,PRS,RSP},    complicating  the prospects of  quantum supremacy  demonstration with Boson Sampling.  Some  of the efficient classical algorithms, due to  imperfections/noise,  are applicable generally, beyond the usual no-collision regime  of Boson Sampling \cite{K1,OB,PRS}. 

   On the other hand, there are necessary and sufficient conditions for classical hardness of imperfect/noisy  Boson Sampling \cite{LP,KK,Arkhipov,VS14,VS15,Brod}.  It is known that an  imperfect  Boson Sampling can be  arbitrarily close in the total variation distance to the ideal one if  the probability of error in the  components of a  network  is  inversely  proportional to the full network depth \cite{AA} times the number of bosons   \cite{Arkhipov} and  distinguishability of  bosons is   inversely proportional to the total number of bosons  \cite{VS15}.          It was shown that  Boson Sampling device with loss of bosons and /or detector noise (dark  counts)   remains as classically hard as the ideal Boson Sampling if the rate of loss of bosons, respectively, that of dark counts, is  inversely proportional to the  number of bosons \cite{Brod}.  It was also suggested that the domain of noise amplitudes allowing  quantum advantage   in a noisy Boson Sampling model  is   scale-dependent,   so that even  noise with  a vanishing amplitude in the total number of bosons leads to vanishing correlations between the output distributions of  noisy Boson Sampling and noiseless one \cite{KK}.  

The above results are found for different models of imperfections/noise in an experimental setup of Boson Sampling. Though sometimes connections were claimed between different models of imperfections,   as in Refs. \cite{KK,Brod,Arkhipov},   precise and rigorous   links  between  all the above results are  still lacking. Moreover, the exact location of the boundary of transition from classical hardness to efficient classical simulations of noisy/imperfect Boson Sampling is not yet known.  The main objective of this work  is  try to fill these two gaps in our knowledge of noisy/imperfect Boson Sampling.    For such a task  we adopt the   Gaussian noise model  of Ref. \cite{KK}, since it has two advantages. First, the model  preserves the main structure of classically hard  mathematical  problem used in Ref. \cite{AA}, namely    approximating  to a multiplicative error    the absolute value of the  matrix permanent of  a matrix of  i.i.d. standard complex Gaussians. Second, as we  reveal in this work,  though being a mathematical abstraction of the effect of noise in Boson Sampling, the model nevertheless has direct links to  experimentally relevant imperfections: losses compensated by dark counts  of detectors (the shuffled bosons model) \cite{Brod} and partial distinguishability of bosons \cite{RR,VS14}. 
We note that our results have other implications. Indeed,  Boson Sampling with  partially distinguishable bosons   can be implemented as a certain   quantum circuit model \cite{CircMod}, which allows to  get insight on classical simulatability of such quantum circuits.

The following (Bachmann–Landau) notations  for relative   scaling  as $N \to \infty$ of two (bounded) functions   $g(N)$ and  $f(N)$    will be used throughout the text: $g= o(f)$   means that $g/ f \to 0$, $g = O(f)$   means that there is such constant $C>0$ that $g/ f  \le C$, $g  = \Omega(f)$   means that there is such constant  $\delta>0$ that $g /f > \delta $,   and finally $g = \omega (f)$ means   $g/ f \to \infty$.   For example, $\Omega(1)$ and $o(1)$ mean, respectively,    bounded from zero (finite)  and vanishing  functions as $N\to \infty$, where as $O(1)$ contains both classes. 

In the next section we state the model of Gaussian noise in Boson Sampling and  recall the main conclusions of Ref. \cite{KK}.  Section   \ref{secOA}     provides an  outline of  the  steps performed in  our  analysis of noisy Boson Sampling model and  states our  main results. Section   \ref{secOB} gives a table containing previous and present results   together  with brief discussion of the results  and relations between them. 
These two sections give an independent summary  of our work, allowing for  quick understanding of   the results,  and help orienting   through  the quite involved technical details of our analysis, presented in the following  sections.    In section \ref{sec3}  the output  probability distribution is derived for the Gaussian noise model of Ref. \cite{KK}. In section \ref{sec4}   the noise model  is extended beyond the no-collision regime. In section \ref{sec5} a connection to Boson Sampling with partially distinguishable bosons is discussed. In section \ref{sec6} we prove  classical simulatability of noisy Boson Sampling with  noise amplitude $ \epsilon = \Omega(1)$. In section \ref{sec7}   we prove classical hardness of the noisy Boson Sampling with  noise  amplitude  $\epsilon=O(1/N)$. The  main results and   open problems are summarised in the concluding section \ref{Concl}.

 \section{The Gaussian noise  model }
 \label{sec2}

We  adopt the    Gaussian noise model in Boson Sampling \cite{KK}, that, on the one hand,   preserves  the mathematical connection to  random Gaussian matrices, used  to  establish   hardness of Boson Sampling  for classical simulations in Ref. \cite{AA}, and, on the other hand,  describes  realistic  sources of noise/imperfections in an experimental implementation  (as we show in this work).  This model of noise,  applicable in the  no-collision regime (i.e., for at least  $M\gg N^2$),   can be  introduced  by   modifying  each element $U_{kl}$ of a unitary network matrix $U$ by   an independent Gaussian  noise, the noisy matrix element $\U_{kl}$ becomes
\be
 \U_{kl}\equiv \sqrt{1-\epsilon} U_{kl} +\sqrt{\epsilon} Z_{kl}
 \en{E1}
where  $Z_{kl}$ is a  rescaled standard complex  Gaussian,     normalised as $\langle |Z_{kl}|^2\rangle =  1/{M}$ and   $\epsilon$ is the  noise amplitude.    Note   Eq. (\ref{E1}) does not mean that we set as   the whole noisy matrix $\U$  a convex combination of   $U$ and  $Z$, since only  submatrices of $\U$ of   size at most $O(N)$ are approximated in this way,  similarly as in  the Gaussian approximation of a unitary network matrix    \cite{AA}.  The crucial difference from Ref. \cite{AA} is  that now  the problem of approximating  the absolute value of  the matrix permanent  is formulated for a  Gaussian matrix  with noisy elements and   efficient classical simulations may be possible depending  on the noise amplitude $\epsilon$.   It was shown  in Ref. \cite{KK} that for  finite noise  amplitude  $\epsilon=  \Omega(1)$  the  noisy Boson Sampling model of Eq. (\ref{E1})  can be  simulated classically with polynomial  in $N$ computations  and that  for  noise amplitude $\epsilon = \omega(1/N)$ the  correlations between   noisy   and ideal Boson Sampling (for   a Haar-random network matrix $U$)  tend to zero. Moreover,   the output distribution of   noisy Boson Sampling    is at a finite total variation distance from that of the ideal Boson Sampling  for  $\epsilon = O(1/N)$, whereas the  classical hardness is retained   for   $\epsilon  = o(1/N)$.    These conclusions were conjectured to hold for  other models of imperfections/noise  and beyond the no-collision regime (i.e., also beyond  the domain of applicability of    Eq. (\ref{E1})). 
  
  Below a  detailed  analysis of the noisy Boson Sampling model of Eq. (\ref{E1}) is attempted in order to establish the threshold of efficient classical simulatability in terms of the noise amplitude $\epsilon$. The crucial step in our analysis is to  establish precise  relation of the Gaussian noise model to other       models of imperfections  allowing to use previous results and methods  to prove  new  results  for noisy Boson Sampling.  The analysis   allows us to get further insight on the location of boundary of efficient  classical simulations  (formulated as a conjecture).   Moreover, we also  find an  extension of the noise model of Eq. (\ref{E1})  beyond the no-collision regime,  where  some of our results preserve validity.  
 
\section{Outline,  results, and connections to previous works} 
\label{Outline}

In this section we give  an outline of our  analysis of the noisy Boson Sampling model of Eq. (\ref{E1}),  section \ref{secOA}, formulate  the main results in the form of two theorems,  and discuss the relations between the present and  previous results on the classical hardness/simulatability of imperfect/noisy Boson Sampling, reformulated for the model of Eq. (\ref{E1}) and presented in a table, section \ref{secOB}. 

\subsection{Outline} 
\label{secOA}
We study   the noisy   Boson Sampling model  of Eq. (\ref{E1})  in the following series of   interrelated steps, illustrated in Fig. 1.  
 
In section \ref{sec3}  we  derive   the output distribution of the noisy Boson Sampling.   This supplies us with a new  (discrete)  representation of the (continuous) Gaussian  noise of Eq. (\ref{E1})  by its effect  on many-boson interference: the noise acts as uniform boson losses, where a  network has uniform transmission $\eta = 1-\epsilon$,   with the lost bosons compensated by  (correlated) random detector clicks at a network output, i.e., we establish  an  equivalence of the  noise model of Eq. (\ref{E1}) and   the shuffled bosons model of Ref. \cite{Brod}.

In section \ref{sec4} the  new    representation of    Gaussian noise model of Eq. (\ref{E1}), found in section \ref{sec3}, is used to extend the latter beyond its domain of applicability, i.e.,  for arbitrary $M\ge N$, beyond the no-collision regime,  by requiring that  the effect of noise on the output probability distribution of Boson Sampling is similar to that in the no-collision regime.  

In section \ref{sec5} we present a model of Boson Sampling with partially distinguishable bosons (and no noise)   which is   equivalent  to the noisy Boson Sampling model  of Eq. (\ref{E1}), if one replaces the  probability factor of dark counts in the discrete   representation of the noisy Boson Sampling of sections \ref{sec3} and \ref{sec4} by that of completely distinguishable bosons (or classical particles), which pass through the network along with indistinguishable bosons.    In section \ref{sec5A}  we argue    that  the  closeness of  the  noisy Boson Sampling model to  the ideal Boson Sampling in the total variation distance~\footnote{The measure of closeness of two probability distributions used in Ref. \cite{AA}.} requires   the noise amplitude   $\epsilon= o(1/N)$ for any $M\ge N$, due to   the results of   Ref. \cite{VS15}. 

In  section \ref{sec6}  we present yet another  equivalent   model of partially distinguishable bosons, with the   same output probability distribution as the model of section \ref{sec5} (i.e., with the same partial distinguishability function \cite{VS14,PartDist}),  and  prove  efficient classical simulatability of the noisy Boson Sampling model in the no-collision regime with finite noise amplitude $\epsilon = \Omega(1)$,   formulated below in theorem 1.   We first prove efficient classical simulatability of the equivalent model with partially distinguishable bosons by utilising  the results of Refs. \cite{VS14,Cliffords,R1,RSP} and then show that   all the steps in the proof  apply  also to the noisy Boson Sampling.     
 
In section \ref{sec7} we find an effective bound on the number of ``noise clicks"  (lost bosons compensated by dark counts of detectors) in the discrete  representation of  the noisy Boson Sampling of section \ref{sec4}, i.e.,  for arbitrary $M\ge N$,       such that Boson Sampling model with bounded total number of  ``noise clicks" is arbitrarily close in the total variation distance to a given   noisy one. This allows us to prove, by utilising the  results of Ref. \cite{Brod}, the  classical  hardness  of the noisy Boson Sampling in the no-collision regime with noise amplitude $\epsilon = O(1/N)$, formulated in theorem 2 below.

 \begin{figure}[ht]
\begin{center}
   \includegraphics[width=.35\textwidth]{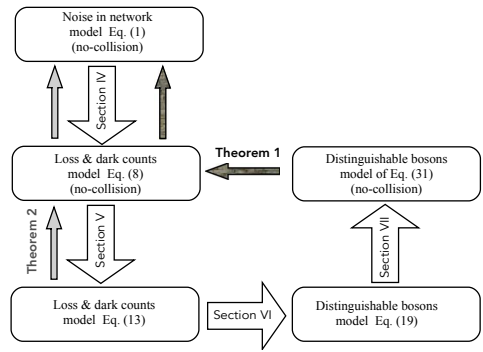}    \caption{Schematic representation of the considered models of imperfect Boson Sampling with indicated equations where they appear and sections which discuss  the models and  relations between them. The models used to prove theorems 1 and 2 (section \ref{secOB} below) are also indicated.   At the bottom of the figure,   the two models are  for  $M\ge N$, beyond the no-collision regime.  \label{F1} }
   \end{center}
\end{figure}
 
\subsection{Main results}
\label{secOB}

The   main results,  proven in the present work in sections \ref{sec6} and \ref{sec7}, respectively, can be stated as follows. 
\begin{theorem} 
Given $0<\delta<1$ and    $\varepsilon>0$, the noisy Boson Sampling model of Eq. (\ref{E1})  in the no-collision regime with  noise amplitude  $\epsilon=  \Omega(1)$    can be  simulated classically, with success probability at least $1-\delta$,   to the   error $\varepsilon$ in the total variation distance  with  computations polynomial $(N,1/\varepsilon,1/\delta)$. 
\end{theorem}
  Theorem 1 agrees with  one of the conclusions  of Ref. \cite{KK}.  On the other hand, we also prove the following. 

\begin{theorem} 
The noisy Boson Sampling model of Eq. (\ref{E1}) with   noise amplitude  $\epsilon= O(1/N)$  is as hard to simulate classically in the  no-collision regime as the ideal Boson Sampling.  
\end{theorem}

 Our results  also indicate  that  the  output distribution of the noisy Boson Sampling of  theorem 2 is at a constant total variation distance to that of the ideal Boson Sampling (as suggested in Ref. \cite{KK}), for arbitrary $M\ge N$, under two plausible conjectures: (i) the noisy Boson Sampling model and the  model of Boson Sampling with partially distinguishable bosons of   section \ref{sec5} are  at the same average  total variation distance from the ideal one   and (ii) that a bound of  Ref. \cite{VS15}  on the  total variation distance to the ideal Boson Sampling  is tight.

Our approach does not resolve if noisy Boson Sampling with the noise amplitude $\epsilon = \omega(1/N)$,  intermediate  between theorems 1 and 2, can be   efficiently simulated classically.  The output distribution of such a noisy Boson Sampling has vanishing correlations with that of the ideal Boson Sampling  \cite{KK}. We show    that  the ratio of the  effective total number of  ``noise clicks"  to the total number of bosons  vanishes as $N\to \infty$ for  noise amplitude  $\epsilon = o(1)$  (valid for arbitrary relation between $N$ and $M$), see  section \ref{sec7}. Such  noise is therefore similar to that with  amplitude $\epsilon=O(1/N)$  by the fact that the dominating contribution  to the output probability distribution comes from the quantum many-boson interferences.  It is natural then to formulate the following conjecture.

\noindent\textbf{Conjecture 1.}\textit{
The noisy Boson Sampling model of Eq. (\ref{E1}) with the noise amplitude  $\epsilon= o(1)$  remains  hard to simulate classically.   
}

The proofs of theorems 1 and 2  are limited  to the no-collision regime,  due to the  limited applicability of the  methods and results of  Refs. \cite{VS14,R1,Brod}, used in the proofs. The main technical  challenges for extension beyond the no-collision regime are  (i)  evaluating the  averages over  the  Haar-random network without using the Gaussian approximation (ii)  extension  of  the quantum computational complexity results to   many-boson interference beyond the no-collision regime.   
 
 \subsection*{Relations with the  previous results} 
\label{relat}
\begin{table}[htbp]
\caption{Reported results on classical hardness/simulatability of imperfect Boson Sampling   reformulated for   noisy Boson Sampling model of Eq. (\ref{E1})  by using the established   equivalence. Here the  ``necessary" and ``sufficient" stand for   the necessary and sufficient bounds on $\epsilon$ for classical hardness of noisy Boson Sampling, whereas ``classical simulations" stands for sufficient condition  for efficient classical simulations.  Results apply  to  no-collision Boson Sampling,  those marked with    ``$\dag$"  apply also   for arbitrary  $M\ge N$.    Bounds  with  ``$*$"   are  conjectures (the question mark denotes possible  interpretation of the result of Ref. \cite{KK}).   }
\vskip 0.25cm
     \resizebox{0.525\textwidth}{!}{\begin{minipage}{1.7\columnwidth}
  \begin{tabular}{|P{1.95cm}|P{1.5cm}|P{1.5cm}|P{2.05cm}|P{2cm}|P{2cm}|}
\hline
  & \multicolumn{3}{|P{4.5cm}|}{ Equivalent  bound on noise  \break amplitude $\epsilon$ }&  \multicolumn{2}{|P{3.5cm}|}{ Bound on  error \break per network  element }  \\
  \hline
 Reference & Necessary   & Sufficient   & Classical simulations &  Necessary    & Sufficient     \\
 \hline
 Ref. \cite{KK}    &     &  $ o\left(\frac{1}{N}\right)$, $\left[o\left(\frac{1}{N}\right)^\dag\right]^*$ &  $\Omega(1)$, $\left[\omega\left(\frac{1}{N}\right)^\dag\right]^* (?)$ &   &   \\
 \hline
 Ref. \cite{LP}   &  $O\left(\frac{\log M}{N^2}\right)$  &   &     & $ O\left(\frac{1}{N^2}\right) $ &    \\
 \hline 
Ref. \cite{Arkhipov}   &      &  $o\left(\frac{1}{N^2}\right)^\dag$ &     &     & $o\left(\frac{1}{N^2\log M}\right)^\dag$  \\
 \hline
Ref. \cite{VS15}   & $ \left[o\left(\frac{1}{N}\right)^\dag\right]^*$    &  $o\left(\frac{1}{N}\right)^\dag$ &     &     &       \\
\hline
Ref. \cite{Brod}   &      &  $O\left(\frac{1}{N}\right)$ &     &     &    \\
   \hline
Refs. \mbox{\cite{R1,RSP}}   &   &      &   $ \Omega(1)$ &   &  \\
   \hline
Refs. \mbox{\cite{OB,PRS}}   &   &      &    $\left[1-o\left(\frac{1}{\sqrt{N}}\right)\right]^\dag $ &   &  \\
  \hline
Present   &   &  $O\left(\frac{1}{N}\right)$, $o(1)^*$  &   $ \Omega(1)$ &   & $O\left(\frac{1}{N\log M}\right)$, $o \left(\frac{1}{\log M}\right)^*$  \\
    \hline
     \end{tabular} 
   \end{minipage}}
  \end{table}

Ref. \cite{RR} mentioned that there may be an  equivalence of partial distinguishability to losses of bosons in their effect on Boson Sampling.  In this work, this is taken further by establishing  actual equivalence of   three noise models, the network noise of Refs. \cite{LP,KK}, the partial distinguishability of bosons \cite{VS14,VS15}, and the shuffled bosons model  of Ref. \cite{Brod}, i.e., when losses of bosons  are compensated by dark counts of detectors.   This equivalence   allows for precise  comparison of  the previous results,  reformulated here for  the noise amplitude $\epsilon$ of Eq. (\ref{E1}), see table 1 (detailed in the discussion below).

   Ref. \cite{LP}   considered a  model of noise in unitary network, where  two-dimensional   unitary matrices $u_i$  from  a product decomposition of a unitary multi-port network $U = \prod_{i}u_i$   \cite{Reck94,Clem2016}    have noisy    elements,  introduced  by the relation  $\Phi(u_i) \Phi(u_i^\dag) = e^{i\tilde\epsilon h}$, where  $\Phi$ is a random  map representing noise,  $\tilde\epsilon>0$ is the amplitude of noise, and  $h$ is a two-dimensional Hermitian matrix consisting of independent     Gaussian  random  variables with zero average and unit variance.  Therefore, the  average probability of an error   per  element is   $O (\tilde\epsilon^2)$, whereas in a random unitary network the total variation distance to the ideal Boson Sampling   becomes  \cite{LP}
 \be
\langle  \mathcal{D} \rangle_{\Phi,U} = \Omega(\tilde\epsilon^2 N^2).
 \en{Ed1} 
To connect the noise  amplitude  $\tilde\epsilon$ of \cite{LP} with  $\epsilon$  of Eq. (\ref{E1}) note that  for a single boson at input the average probability of an error to occur is $O(\tilde\epsilon^2 d)$, where $d$ is the network depth   $d= O(\log M)$    \cite{AA} (i.e., the current  necessary number of elementary blocks in the unitary   matrix factorisation  by $2$-dimensional ones,  as in Ref.    \cite{Reck94}). On the other hand,  in the Gaussian noise model Eq. (\ref{E1})  the average  probability of an error is $O(\epsilon)$ (as it is  quadratic  in the matrix element). We have $\epsilon \sim \tilde \epsilon^2 \log M$  (first  established  in    Ref. \cite{Arkhipov}).   Therefore, the necessary  condition  on  the average probability of error per network  element    $\tilde\epsilon^2 = O(1/N^2)$ \cite{LP}, for classical hardness of such a noisy Boson Sampling,    in our terms  reads  $\epsilon=  O( \log M/N^2)$.  

On the other hand, a sufficient condition on imperfect/noisy unitary matrix $\tilde{U}$  was obtained in Ref. \cite{Arkhipov} by establishing  a bound on the total variation distance 
 \be
 \mathcal{D} \le N || U- \tilde{U}||,
 \en{Ed4}
 where $|| \ldots ||$ is the matrix norm.  There it was also shown that  for the Gaussian noise model  of Ref. \cite{KK}  
 \be
 ||U - \U || = O(\sqrt{\epsilon}).
 \en{Ed5}
By  Eqs. (\ref{Ed4})-(\ref{Ed5}) a sufficient condition for  a small total variation distance $\mathcal{D} = o(1)$  to the  ideal Boson Sampling   reads $\epsilon  = o (1/N^2)$ \cite{Arkhipov}.  Theorem 2 reduces the sufficient condition for classical hardness to $\epsilon = O(1/N)$. Therefore, it  relaxes   also the sufficient condition on the    average probability of    error per network element  from  $o(1/(N^2\log M))$  (according to the  above result of  Ref. \cite{Arkhipov}) to   $O(1/(N\log M))$. Note that this new sufficient bound does not satisfy the necessary one of Ref. \cite{LP} (the resolution of this apparent contradiction  is given  below).

 Ref. \cite{VS15}  gives a sufficient bound    on boson distinguishability,  i.e., on the overlap of internal states of bosons $\xi = 1-o(1/N)$,    sufficient  for the total variation distance to the ideal Boson Sampling to be $\mathcal{D} = o(1)$ for arbitrary $M\ge N$.  This condition corresponds  to the noise amplitude $\epsilon = o(1/N)$ (under the conjecture that the two models of noise are equivalent, see sections \ref{sec5A} and \ref{sec6}), which  agrees with the  noise stability of Boson Sampling shown in Ref. \cite{KK} for  $\epsilon = o(1/N)$.    Moreover, the condition $\epsilon = o(1/N)$  is conjectured in Ref. \cite{VS15} to be  also  necessary for vanishing total variation distance   to the ideal Boson Sampling (again, there is an apparent contradiction with the sufficient bound due to theorem 2,  $\epsilon = O(1/N)$, see below). 
 
 We   use the  equivalence between the models of imperfections   and the results of Ref. \cite{Brod} to prove theorem 2 stating  that noisy Boson Sampling with noise amplitude $\epsilon = O(1/N)$ is as classically hard as the ideal Boson Sampling.   Ref. \cite{KK} (and the above discussed conjecture of Ref. \cite{VS15}) states   that  such  noisy   Boson Sampling  is at a constant total variation distance from the ideal one. These facts give the resolution of  the apparent  disagreement  between theorem 2 and the necessary bounds of  Refs.  \cite{LP,KK,VS15}:    closeness to  the  ideal  (noiseless)  Boson Sampling is  \textit{not the only  possible} reason  for classical  hardness  of noisy Boson Sampling.    We can conclude that  the necessary bounds on noise/imperfection amplitude derived before  by requiring a small total variation distance to the ideal Boson Sampling may be  actually not  necessary for the classical hardness of noisy/imperfect Boson Sampling.   This   also makes    the    analysis   of Ref. \cite{BSECT}   irrelevant, since the  classical hardness of  an imperfect/noisy realisation of  Boson Sampling  was assumed  there  to be  equivalent to it being at  a small   total variation distance from the ideal model.

Theorem 1 states efficient  classical simulatability of  no-collision Boson Sampling on a  noisy network  with the noise  amplitude $\epsilon =  \Omega(1)$, additionally to the previously proven classical simulatability for finite partial distinguishability, i.e., finite overlap of internal states of bosons, $\xi = 1- \Omega(1)$,   \cite{R1}, and for    losses of bosons proportional to the total number of them   \cite{RSP}, i.e., with the transmission $\eta = 1-\Omega(1)$.

In Refs. \cite{OB,PRS} it was shown that Boson Sampling on a lossy network with the  (maximal) transmission $\eta = o(1/\sqrt{N})$  can be efficiently simulated classically for any $M\ge N$. This  bound  transforms into an equivalent one  on the amplitude of noise in Eq. (\ref{E1}) $\epsilon = 1 - o(1/\sqrt{N})$. Ref. \cite{K1}  has  shown that lossy Boson Sampling with any finite transmission and   stronger dark counts than in the  discrete representation of noisy Boson Sampling, derived below in  section \ref{sec3},  can be efficiently simulated classically. However, this result does not apply beyond the no-collision regime, since it depends on the usage of boson number  unresolving detectors,  which are sufficient only in the no-collision regime \cite{AA}.

Finally,  if conjecture 1  of section \ref{secOA} on   classical hardness  of noisy Boson Sampling with   the noise with amplitude $\epsilon=o(1)$   is true, this would mean that   the average acceptable probability of error per network element has to be just    $o(1/\log M)$, i.e., smaller than  the  inverse of      network depth \cite{AA}.   
   
 \section{Output probability distribution of noisy Boson Sampling}
\label{sec3}
Let us find  the output probability distribution of the noisy Boson Sampling model  in  the no-collision regime, given by Eq. (\ref{E1}).  We fix the input ports to  be $k = 1,\ldots,N$.   Assuming  that the  matrix  $Z$   changes from run to run,  let us derive the  probability  $ \langle p^{(\U)}(l_1\ldots l_N|1\ldots N)\rangle$ to detect $N$ input  bosons   at distinct output ports $l_1,\ldots,l_N$.    Denoting by $\mathcal{M}(k_1\ldots k_n|l_1 \ldots l_n)$ the submatrix of a matrix $\mathcal{M}$ on the rows $k_1,\ldots,k_n$ and columns $l_1, \ldots, l_n$, by $\per\mathcal{M}(k_1\ldots k_n|l_1 \ldots l_n)$ the matrix permanent of such a submatrix, and  (in this and the next section)  by $\langle \ldots \rangle$   the averaging   over the noise $Z$,  we have  \begin{widetext}
\begin{eqnarray}
\label{E2}
&& \langle p^{(\U)}(l_1\ldots l_N|1\ldots N)\rangle  = \left\langle \bigl| \per \U(1\ldots N| l_1 \ldots l_N)  \bigr|^2\right\rangle \nonumber\\
& &=\left\langle\Biggl| \sum_{n=0}^N \epsilon^{\frac{n}{2}} (1-\epsilon)^{\frac{N-n}{2}}\sum_{k_1\ldots k_n} \sum_{{j_1}\ldots {j_n}} \per Z(k_1\ldots k_n|l_{j_1}\ldots l_{j_n})\per U(k_{n+1}\ldots k_N|l_{j_{n+1}}\ldots l_{j_N})\Biggr|^2\right\rangle,
\end {eqnarray}
\end{widetext}
where  we partition the set $1,\ldots,N$ twice into  two groups of  subsets:  (i) $k_1,\ldots,k_n$ and $ k_{n+1},\ldots,,k_N$ and (ii)   ${j_1},\ldots ,{j_n}$ and ${j_{n+1}},\ldots ,{j_N}$ (here and below the notation  $\sum_{k_1\ldots k_n}$    stands for the summation over all subsets, $k_1,\ldots,k_n$,  taken from a larger set, in Eq. (\ref{E2}) from $1,\ldots,  N$) and expand   the matrix permanent of the  sum of two matrices in Eq. (\ref{E1})  using  the permanent expansion formula \cite{Minc}. The averaging over the noise is  easily performed within the expansion in Eq. (\ref{E2}), where we need to consider the averaging of the   products of the following type (we use  that    each output probability involves only a submatrix of $\U$ of size $N$, therefore we are in the  applicability of domain of approximation by Eq. (\ref{E1}), see the note below Eq. (\ref{E1}) in section \ref{sec2})
\begin{eqnarray}
\label{E3}
&&\langle \per  Z(k_1\ldots k_n|l_{j_1}\ldots l_{j_n}) \per  Z^*(k^\prime_1\ldots k^\prime_{m}|l_{j^\prime_1}\ldots l_{j^\prime_{m}})\rangle \nonumber\\
&& = \sum_{\sigma\in S_n} \sum_{\tau\in S_m} \left\langle 
\left[\prod_{\alpha=1}^n Z_{k_{\sigma(\alpha)},l_{j_\alpha}} \right]
\prod_{\beta=1}^m Z^*_{k^\prime_{\tau(\beta)},l_{j_\beta^\prime} }
\right\rangle\nonumber\\
&& = \frac{\delta_{n,m}}{M^n} \sum_{\sigma,\tau\in S_n}\prod_{\alpha=1}^n  \delta_{k_{\sigma(\alpha)},k^\prime_{\tau(\alpha)}}\delta_{l_{j_\alpha},l_{j^\prime_\alpha}}\nonumber\\
&& = \delta_{n,m} \frac{n!}{M^n} \prod_{\alpha=1}^n \delta_{k_{\alpha},k^\prime_{\alpha}}\delta_{l_{j_\alpha},l_{j^\prime_\alpha}},
\end{eqnarray}
where we have used  that $Z$ is a matrix of i.i.d. Gaussians with $\langle Z_{kl}\rangle = 0$ and $\langle| Z_{kl}|^2\rangle = 1/M$. 
Inserting the result of Eq. (\ref{E3}) into Eq. (\ref{E2}),   introducing the  binomial distribution 
\be
B_n(x) \equiv \binom{N}{n} x^n (1-x)^{N-n}
\en{E4}  
and rearranging the summations (with  $n$ and  $N-n$ interchanged) we obtain  the output  probability of Eq. (\ref{E2})   as follows
\begin{eqnarray}
\label{E5}
 &&\langle p^{(\U)}(l_1\ldots l_N|1\ldots N)\rangle = \sum_{n=0}^N B_n(1-\epsilon)    \frac{(N-n)!}{M^{N-n}}\nonumber\\
 && \times \binom{N}{n}^{-1}  \sum_{{j_1}\ldots {j_n}} \sum_{k_1\ldots k_n}   p^{(U)}_q(l_{j_1}\ldots l_{j_n}|k_1\ldots k_n),
 \end{eqnarray}
where 
\be
p^{(U)}_q(l_{j_1}\ldots l_{j_n}|k_1\ldots k_n) = \left|\per U(k_1\ldots k_n|l_{j_1}\ldots l_{j_n})\right|^2.
\en{E6}
Eq. (\ref{E6}) gives   the probability  of  $n$ indistinguishable bosons,   sent to  distinct  input ports $k_1,\ldots,k_n$  of the unitary network $U$, to be detected  at  distinct output ports $l_{j_1},\ldots ,l_{j_n}$ \cite{C}.

All the factors on the r.h.s. of Eq. (\ref{E5}) have  physical interpretation as some probabilities.  The   factor  $B_n(1-\epsilon)$  is the probability that only $n$ of the total $N$ input bosons, from $n$ random input ports,   remain indistinguishable during the propagation in the noisy network. These bosons  contribute the probability factor  
 \be
 \overline{ p^{(U)}_q}(l_{j_1}\ldots l_{j_n}) 
 \equiv \binom{N}{n}^{-1} \sum_{k_1\ldots k_n}p^{(U)}_q(l_{j_1}\ldots l_{j_n}|k_1\ldots k_n).
\en{E7} 
 The rest $N-n$ of the input bosons behave as   distinguishable  bosons (or classical particles),  uniformly randomly populating  the output ports, thus contributing the  probability  factor $(N-n)!/M^{N-n}$\footnote{In the no-collision regime with the probability \mbox{$1- O(N^2/M)$} they populate  distinct output ports $l_{n+1},\ldots, l_{N}$ that complement the distinct  ports $l_1,\ldots, l_n$ where the  indistinguishable bosons end up.}.   
Observe that the Gaussian  noise in Boson Sampling   is now represented as    uniform losses  with the transmission $1-\epsilon$ (Eq. (\ref{E7})  is equivalent to that for a  Boson Sampling with a lossy input, with   only $n$ out of $N$ bosons making to a network),   compensated by  dark  (random) counts of detectors at  output ports:  there so many (uniformly distributed)  dark counts   as  the lost bosons, so that the total number of   detector clicks  is equal to the total number of input bosons. The action of noise is therefore similar to the model of  shuffled  bosons  discussed in Ref. \cite{Brod} (note, however,  that the probability distribution of our ``dark counts"   is different from that of  physical  dark counts of detectors,  see more on this in  section \ref{sec4}).  We will use   this equivalence in section \ref{sec7} to prove theorem 2 of section \ref{secOB}.

\section{Extending the Gaussian noise  model beyond the no-collision regime}
\label{sec4}
The discrete representation of the Gaussian noise, Eq. (\ref{E5}) of  section \ref{sec3}, allows one to  extend the noisy Boson Sampling  model of Ref. \cite{KK} for arbitrary $N,M$, i.e.,  beyond the usual  no-collision regime. Generally speaking, such an extension is not unique. We choose  an extension of noise  that has  a similar   effect on the  many-boson  interference in the general case as the Gaussian noise   in the no-collision regime.  Since all the factors in Eq. (\ref{E5}) are interpreted as some probabilities, such an extension  can be easily obtained by generalising them  and the summation over  distinct output ports $l_{j_1},\ldots, l_{j_n}$ in Eq. (\ref{E5}) for general  $N,M$. 

For arbitrary $N,M$ there are multiply occupied  output ports,   hence, we have a multi-set $l_1\le \ldots \le l_N$. Let us fix the notations for below. We denote by $\m= (m_1,\ldots,m_M)$, $|\m|\equiv m_1+\ldots + m_M=N$,  the output configuration corresponding to a multi-set $l_1\le \ldots \le l_N$ (where $m_l$ is the total number of bosons in output port $l$), whereas  $\s = (s_1,\ldots,s_M)$,   $|\s| = n$,  and $\br = (r_1,\ldots,r_M)$,  $|\br|=N-n$,  will denote the output  configurations corresponding, respectively,  to the   multi-subsets  $l_{j_1}\le \ldots \le l_{j_n}$ and  $l_{j_{n+1}}\le \ldots \le l_{j_N}$ of the above multi-set (thus $\m = \br +\s$). We also set $\m! \equiv m_1!\ldots m_M!$. 

First, the probability of  an output  configuration  $\br$   obtained by  uniformly  randomly distributing   $N-n$   distinguishable bosons,   or,  equivalently  \cite{PartDist}, indistinguishable classical  particles   over the output ports   reads 
\be
q(\br) =  \frac{(N-n)!}{\br! M^{N-n}}.
\en{Cprob}
 Eq. (\ref{Cprob}) can be derived as follows. Assume that  classical particles are enumerated (i.e., classical and distinguishable), then the probability to get any given   output  is $1/M^{N-n}$, since   each time one chooses   one of $M$ ports uniformly randomly.  Ignoring the   identities of the particles,  we obtain for  an output configuration $\br$   exactly $(N-n)!/\br!$ distributions of distinguishable classical particles, i.e., the probability in Eq. (\ref{Cprob}).

Second, the   summation  over the  subsets of \textit{distinct} output ports $l_{j_1}< \ldots< l_{j_n}$,  (equivalently,  over the    subindices $1\le {j_{1}}<  \ldots < {j_n}\le N$) in Eq. (\ref{E5})   must be  replaced for  general $M\ge N$ by  the  summation over all  sub-configurations  $\s\subset \m$ (i.e., $s_\alpha \le m_\alpha$ for all $\alpha =1,\ldots, M$),  each corresponding to a multi-set of the output ports $l_{j_{1}}\le \ldots \le  l_{j_n}$, to ensure that we do not double count  the same output configurations (since different subsets  $1\le {j_{1}}<  \ldots < {j_n}\le N$ may correspond to the same output configuration $\s$). There is a  mathematical relation between the two types of summations valid for any  symmetric function $f(l_1,\ldots,l_N)$ (e.g., Ref.  \cite{Minc})
\be
\sum_{{\s \subset\m \atop |\s|=n}} f(l_1,\ldots,l_N)= \sum_{{j_1}\ldots {j_n}} f(l_1,\ldots,l_N)\prod_{\alpha=1}^M \binom{m_\alpha}{s_\alpha}^{-1}.
\en{E8}

 With  the above two observations,   replacing   the output port indices by the corresponding occupations, i.e.,  $l_1\ldots  l_N$ by $\m$ and $l_{j_1}\ldots  l_{j_n}$ by $\s$,   the probability distribution of Eq. (\ref{E5}) generalised beyond the no-collision regime becomes 
\begin{eqnarray}
\label{E9}
 && \langle p^{(\U)}(\m |1\!\ldots\! N)\rangle \equiv \sum_{n=0}^N B_n(1-\epsilon)  \nonumber\\
 && \times \binom{N}{n}^{-1} \sum_{k_1\ldots k_n}   \sum_{{\s \subset\m \atop |\s|=n}}  q(\br) p^{(U)}_q(\s |k_1\!\ldots\! k_n),\nonumber\\
\end{eqnarray}
here  $p^{(U)}(\s|k_1\ldots k_n) $ is   the probability for indistinguishable bosons from input ports $k_1,\ldots, k_n$  to be detected in  output configuration  $\s$ corresponding to multi-set of output ports $l_{j_1}\le \ldots \le l_{j_n}$  \cite{C,Scheel,AA}
\be
p^{(U)}_q(\s|k_1\!\ldots\! k_n) = \frac{1}{\s!} \left|\per U(k_1\ldots k_n|l_{j_1}\ldots l_{j_n})\right|^2,
\en{E10}
generalising  that of Eq. (\ref{E6}).  One can verify by straightforward calculation with  use of the  summation  identity \cite{Minc}, 
\be
\sum_{|\m|=N} f(l_1,\ldots,l_N) = \sum_{l_1=1}^M \dots \sum_{l_N=1}^M \frac{\m!}{N!}f(l_1,\ldots,l_N), 
\en{IdSum}
  valid for any symmetric function $f(l_1,\ldots,l_N)$, and unitarity of the network matrix $U$,  that the probabilities in Eq. (\ref{E9}) sum to $1$,
\[
\sum_{|\m|=N} \langle p^{(\U)}(\m|1\ldots N)\rangle = 1.
\]

The equivalent representation of the Gaussian  noise found in section \ref{sec3}, Eq. (\ref{E5}), and   extended here for   $M\ge N$, Eq. (\ref{E9}),  is the basis for our analysis of the effect of noise on Boson Sampling.   

Let us return here to the equivalence of the noisy Boson Sampling model Eq. (\ref{E9}) to that with losses and dark counts of detectors, discussed  at the end of section \ref{sec3}.  There is some difference between our model of dark counts and the  dark counts of \textit{independent} physical detectors,  with each detector following Poisson distribution $p(n) = \frac{\nu^n}{n!} e^{-\nu}$,  $\nu>0$. The probability of $N-n$    dark counts  of physical detectors in an output configuration $\br$ would be in this case given by the distribution 
\be
\tilde{q}(\br) =  \frac{\nu^{N-n}}{\br!} e^{-M\nu},
\en{PDC}
which is different from the probability $q(\br)$ of Eq. (\ref{Cprob}) describing a model of   \textit{correlated dark counts} at output ports.   The output distribution in  Eq. (\ref{E9})  shows nevertheless  equivalence to the shuffled bosons model of Ref. \cite{Brod},  where  the  lost bosons are  compensated exactly by the dark   counts of detectors in uniformly  random output ports. Moreover, as we discuss below, there is also an  approximate equivalence of the noisy Boson Sampling model to that with partially distinguishable bosons (and no noise).

\section{Noise in Boson Sampling and   partial distinguishability of bosons}
\label{sec5}
 
The  representation of the  Gaussian noise  by its action on  many-boson interference, discussed in sections \ref{sec3} and \ref{sec4},   allows one also to reveal a relation between   the effect of  noise   and   that  of partial distinguishability of bosons on  classical hardness of  Boson Sampling.  Such a relation will  be useful  in the  proof of  classical simulatability of the noisy BosonSamling model (section\ref{sec6}). 

In the discrete  representation of  noisy Boson Sampling model,  Eq. (\ref{E9}),   distinguishable bosons  are uniformly randomly distributed at the output ports (equivalent of the  correlated dark detector counts) to compensate for the lost indistinguishable bosons.   Consider a related imperfect Boson Sampling  model, where $N-n$ distinguishable bosons   are also sent through a  unitary (noiseless) network along with $n$ indistinguishable ones, complementing the input ports to $1,\ldots, N$,  with the total number $n$ of the  indistinguishable bosons  distributed  according to the binomial of  Eq. (\ref{E4}) with $x = 1-\epsilon$.   The  probability of uniformly randomly distributing $N-n$ particles over the output ports  of Eq. (\ref{Cprob}) must be   replaced  in this case   by the   classical probability of the respective  output sub-configuration $\mathbf{r}$ of $N-n$ distinguishable bosons (which depends on the absolute values squared of the  network matrix elements).   We conjecture that  replacing  in Eq. (\ref{E9})  the uniform probability with a  classical probability  has no effect on  the classical hardness of the  output distribution. Indeed, classical probabilities are  given by the matrix permanents of doubly-stochastic matrices (with the elements $|U_{kl}|^2$),  simulated by  an  efficient classical algorithm   \cite{JSV}. The conjecture is  also  partially  confirmed below: (i) by comparison of   our condition on the  noise amplitude, obtained under the conjecture in subsection \ref{sec5A},  with the previously established bounds \cite{LP,Arkhipov}  and (ii)    in   section \ref{sec6}, where we find  that conditions on classical simulatability of the Boson Sampling with partially distinguishable bosons and that  on  the noisy Boson Sampling   are the same for the same amplitude of respective  imperfections. 

 Before continuing, let us  recall the basics of the partial distinguishability theory  (see Refs. \cite{VS14,PartDist,Ninter} for  more details).  In this theory, bosonic degrees of freedom are partitioned into two types: the operating modes and the internal states.  A unitary network  performs a unitary transformation  $U$ of  the operating modes, whereas the internal modes allow to distinguish bosons, at least in principle. Assuming that bosons are in the same  \textit{pure} internal state (i.e., the state over all other degrees of freedom other than the operating modes \cite{PartDist})  one obtains  the output probability formula of Eq. (\ref{E10}) \cite{C,AA}.  The usual  operating modes of single photons, for example, are the    propagating modes of a spatial  optical network. Another possibility is to use the  temporal operating  modes, i.e., the  time-bins \cite{TbinBS,TbinBSExp}.  
However,    $N$ single  photons, usually coming from distinct sources or at different times from a single source, are  not  in the same internal  pure state.  Single photon  generation is a probabilistic process   with all the available to-date  sources (see for instance Ref.  \cite{SPS}),   experimentally available single    photons  are in some mixed states.  Assuming that  the internal states are not  resolved (such resolution only adds network-independent randomness to boson counts), the   probability  of an output configuration $\m=(m_1,\ldots,m_M)$ (corresponding to a multi-set of  output ports $l_1\le \ldots \le l_N$)  in   the general  case of    partially distinguishable   bosons at the input ports $k=1,\ldots, N$  becomes   \cite{VS14,PartDist,Ninter}
\be 
p^{(U)}_J(\m|1\!\ldots\! N) \!= \!\frac{1}{\m!}\sum_{\sigma_1,\sigma_2}\! J(\sigma_1\sigma^{-1}_2) \prod_{k=1}^N U^*_{\sigma_1(k),l_k}U_{\sigma_2(k),l_k}, 
\en{E15}
where the double summation  with   $\sigma_1,\sigma_2$ runs  over the symmetric group $S_N$ of $N$ objects. Here a complex-valued (distinguishability) function $J(\sigma)$   depends on the internal state of $N$ bosons $\rho\in H^{\otimes N}$, where $H$ is the internal Hilbert space of single boson,  and is defined as follows \cite{VS14,PartDist}
\be
J(\sigma) = \mathrm{Tr}\{P_\sigma \rho \}, 
\en{E16} 
with $P_\sigma$ being the unitary operator representation of the symmetric group $S_N$ action in the tensor product 
$H^{\otimes N}$. 
In the case of completely indistinguishable bosons $J(\sigma)=1$,  for all $\sigma\in S_N$,  we recover from Eqs. (\ref{E15})-(\ref{E16}) the probability of Eq. (\ref{E10}), whereas for $J(\sigma) = \delta_{\sigma,I}$,  with $I$ being the identity   permutation  (orthogonal internal states), the probability formula for completely distinguishable bosons (or indistinguishable classical particles) \cite{VS14,PartDist}. 

The discussion at the beginning of this section can be summarised by replacing    the output distribution of Eq. (\ref{E9}) by  the following  distribution
\begin{eqnarray}
\label{E11}
  && p^{(U)}_J(\m| 1\ldots N) = \sum_{n=0}^N B_n(1-\epsilon) \binom{N}{n}^{-1}   \\
  & &\times  \sum_{k_1\ldots k_n}   \sum_{{\s\subset\m \atop |\s|=n}} p^{(U)}_q(\s |k_1\!\ldots\! k_n) p^{(U)}_{cl}(\br|k_{n+1}\ldots k_N),\nonumber
 \end{eqnarray}
here the probability   factor (\ref{Cprob})  in Eq. (\ref{E9}) is replaced by the  classical probability  of    $N-n$ distinguishable bosons, sent at  the  inputs $k_{n+1},\ldots,k_N$,   to be detected in the   configuration $\br$ of   output ports, i.e., 
\begin{eqnarray}
\label{E12}
&& \!\!\!\!\!\!\!  p^{(U)}_{cl}(\br |k_{n+1}\!\ldots\! k_N)\!=\!\frac{1}{\br!} \sum_{\sigma\in S_{N\!-\!n}} \prod_{\alpha=n\!+\!1}^{N} |U_{k_{\sigma(\alpha)},l_{j_\alpha}}|^2 \nonumber\\
&&  = \frac{1}{\br!}\per |U|^2(k_{n+1}\!\ldots\! k_N| l_{j_{n+1}}\!\ldots \! l_{j_N}),
\end{eqnarray}
where $l_{j_{n+1}}\le \ldots\le  l_{j_N}$ are the output ports corresponding to $\br$. 
For instance,  in a uniform network $|U_{kl}| = \frac{1}{\sqrt{M}} $  the  probabilities of Eqs. (\ref{Cprob}) and (\ref{E12}) coincide.

 Now, we can introduce a model of partial distinguishability which results precisely in  the output distribution of Eq. (\ref{E11}) (another such   model is presented   in section \ref{sec6}). Consider $N$ single bosons, where boson  at input port $k$ is  in the following internal state 
\be
\rho_k = (1-\epsilon) |\phi_0\rangle\langle \phi_0| + \epsilon |\phi_k\rangle\langle \phi_k|, 
\en{E17} 
where  $\langle \phi_i|\phi_j\rangle  = \delta_{i,j}$ for all $i,j\in \{0,1,\ldots,N\}$.  Eq. (\ref{E17}) means that  with probability $1-\epsilon$ boson   $k$ is  in the common pure state $|\phi_0\rangle$ and with probability $\epsilon$ in a unique  ortogonal state $|\phi_k\rangle$.  The internal state of $N$ such  independent  bosons  reads $\rho = \rho_1\otimes \ldots \otimes  \rho_N$.  By Eq. (\ref{E17}) and orthogonality of the specific states $|\phi_k\rangle$,   the probability to have  $n$ indistinguishable single bosons  (at randomly chosen  input ports) is $B_n(1-\epsilon)$, where $B_n(x)$ is the binomial distribution (\ref{E4}).    The partial distinguishability  function  is straightforward to obtain from the definition (\ref{E16}). By expanding the tensor product $\rho_1\otimes \ldots \otimes  \rho_N$ using the expression of Eq. (\ref{E17}) and substituting the result into Eq. (\ref{E16}) we get 
 \begin{eqnarray}
 \label{E19} 
&&J(\sigma) = \sum_{n=0}^N (1-\epsilon)^n\epsilon^{N-n} \sum_{k_1\ldots k_n} \prod_{\alpha=n+1}^N \delta_{k_{\sigma(\alpha)},k_\alpha} \nonumber\\
&& = \sum_{n=0}^N (1-\epsilon)^n\epsilon^{N-n} \sum_{k_1\ldots k_n}  \sum_{\tau\in S_n}\delta_{\sigma,\tau\otimes I} ,
\end{eqnarray}
where  the two subsets $k_1,\ldots,k_n$ and $k_{n+1},\ldots,k_N$  of $1,\ldots,N$ give  the contributions from the first and the second terms in Eq. (\ref{E17}), respectively,  (the second subset contains only the  fixed points of permutation $\sigma$ by orthogonality of the unique states $|\phi_k\rangle$, $k=1,\ldots,N$). Thus for each subset $k_1,\ldots,k_n$ the  nonzero terms in Eq. (\ref{E19}) correspond to permutations of the type $\sigma = \tau\otimes I$, with $\tau\in S_n$ acting on this subset. 

Using the partial distinguishability of Eq. (\ref{E19}) in  the general  formula (\ref{E15}) reproduces the result of Eq. (\ref{E11}). Indeed, modifying slightly Eq. (\ref{E15}), by setting $\sigma = \sigma^{-1}_2$ and $\sigma_R = \sigma_1\sigma_2^{-1}$, reordering the factors in the  product,  and using  the identity of Eq. (\ref{E8}) we get
\begin{widetext}
\begin{eqnarray*}
&& p^{(U)}_J(\m |1\ldots N) = \frac{1}{\m!} \sum_{\sigma_R,\sigma\in S_N} J(\sigma_R)\prod_{k=1}^N U^*_{\sigma_R(k),l_{\sigma(k)}} U_{k,l_{\sigma(k)}} \nonumber\\
&& =  \frac{1}{\m!} \sum_{n=0}^N B_n(1-\epsilon) \binom{N}{n}^{-1}  \sum_{k_1\ldots k_n} \sum_{\tau_R\in S_n}  \underbrace{\sum_{\sigma\in S_N} }\left[ \prod_{\alpha=1}^n U^*_{k_{\tau_R(\alpha)},l_{\sigma(k_\alpha)}} U_{k_\alpha,l_{\sigma(k_\alpha)}}\right] \prod_{\alpha=n+1}^N |U_{k_\alpha,l_{\sigma(k_\alpha)}}|^2
\nonumber\\
&& =  \frac{1}{\m!} \sum_{n=0}^N B_n(1-\epsilon) \binom{N}{n}^{-1}  \sum_{k_1\ldots k_n}\sum_{\tau_R\in S_n} \underbrace{\sum_{j_1\ldots j_n} \sum_{\pi\in S_n}   \sum_{\mu\in S_{N-n}} } \left[ \prod_{\alpha=1}^n U^*_{k_{\tau_R(\alpha)},l_{j_\pi(\alpha)}} U_{k_\alpha,l_{j_\pi(\alpha)}}\right] \prod_{\alpha=n+1}^N 
|U_{k_\alpha,l_{j_{\mu(\alpha)}}}|^2
\nonumber\\
&& =  \frac{1}{\m!}  \sum_{n=0}^N B_n(1-\epsilon) \binom{N}{n}^{-1}  \sum_{k_1\ldots k_n}\sum_{j_1\ldots j_n} \left[\sum_{\nu\in S_n}     \prod_{\alpha=1}^n U^*_{k_{\alpha},l_{j_{\nu(\alpha)}}} \right]\left[\sum_{\pi\in S_n}  \prod_{\alpha=1}^n U_{k_\alpha,l_{j_\pi(\alpha)}}\right] \sum_{\mu\in S_{N-n}}  \prod_{\alpha=n+1}^N
 |U_{k_\alpha,l_{j_{\mu(\alpha)}}}|^2
 \nonumber\\
&& = \sum_{n=0}^N B_n(1-\epsilon) \binom{N}{n}^{-1}  \sum_{k_1\ldots k_n}\sum_{j_1\ldots j_n}   \left[ \prod_{\alpha=1}^M \binom{m_\alpha}{s_\alpha}^{-1} \right] p^{(U)}_q(\s |k_1\ldots k_n) p^{(U)}_{cl}(\br |k_{n+1}\ldots k_N)\nonumber\\
&&= \sum_{n=0}^N B_n(1-\epsilon) \binom{N}{n}^{-1}  \sum_{k_1\ldots k_n}\sum_{{\s\subset\m \atop |\s|=n}}   p^{(U)}_q(\s |k_1\ldots k_n) p^{(U)}_{cl}(\br |k_{n+1}\ldots k_N),
\end{eqnarray*}
\end{widetext}
where  we have taken into account that by Eq. (\ref{E19}) $\sigma_R = \tau_R\otimes I$,   introduced $\nu = \pi \tau_R^{-1}$ to factor the summations over the permutations,  and used the underbraces  to show the factorisation of  permutation $\sigma\in S_N$  as follows   $\sigma = (\pi\otimes \mu)\rho$ with arbitrary   permutations $\pi\in S_n$ and $\mu\in S_{N-n}$ and  $\rho\in S_N/(S_n\otimes S_{N-n})$  (i.e., from the factor group) selecting a subset $j_1,\ldots, j_n$ from $1,\ldots, N$.

\subsection{Sufficient condition on  noise  amplitude for closeness of noisy Boson Sampling  and the ideal one }
\label{sec5A}

In Ref. \cite{VS15} a plausibly tight bound was found for closeness in the total variation distance  of the output distribution of  imperfect Boson Sampling with partially distinguishable bosons to that of the ideal Boson Sampling. Here we consider the bound  in relation to the noisy Boson Sampling model. The bound is  as follows. The output probability distribution   of Boson Sampling  with  $N$ partially distinguishable bosons  in an internal state given by $J(\sigma)$ (\ref{E16}) (and no other imperfections in the setup)  is at most at $(1-d_J(N))$-distance   in the total variation distance to the  ideal Boson Sampling, 
\be
\mathcal{D}  = \frac{1}{2}\sum_{|\m|=N} |p^{(U)}_q(\m) - p^{(U)}_J(\m)|\le 1-d_J(N),
\en{E20}
where
\be
d_J(N) =  \frac{1}{N!}\sum_{\sigma\in S_N}  \mathrm{Tr}\{P_\sigma \rho^{(int)} \}= \frac{1}{N!}\sum_{\sigma\in S_N}J(\sigma).
\en{E21}
The bound of Eq. (\ref{E20}) can be easily understood: the quantity $d_J(N)$ of  Eq. (\ref{E21}) is the  magnitude  of  projection on the  subspace of  $H^{\otimes N}$ consisting of  the completely symmetric internal states (such an internal state corresponds to the completely indistinguishable bosons \cite{PartDist}). Indeed, the projection  reads
\be
\rho^{(S)} = \hat{S}_N \rho \hat{S}_N,\quad \hat{S}_N \equiv \frac{1}{N!}\sum_{\sigma\in S_N} P_\sigma,
\en{E22}
giving $d_J(N) =    \mathrm{Tr}\{\rho^{(S)}\} = \mathrm{Tr}\{\hat{S}_N \rho\}$ by the fact that $\hat{S}^2_N = \hat{S}_N$. We obtain from Eqs. (\ref{E19}) and  (\ref{E21})  
\begin{eqnarray}
\label{E23}
&& d_J(N) = \frac{1}{N!} \sum_{n=0}^N \epsilon^n(1-\epsilon)^{N-n}\sum_{k_1\ldots k_n} \sum_{\sigma\in S_N}\prod_{\alpha=n+1}^N\delta_{k_{\sigma(\alpha)},k_\alpha} \nonumber\\
&& = \frac{1}{N!} \sum_{n=0}^N \epsilon^n(1-\epsilon)^{N-n}\sum_{k_1\ldots k_n}  \sum_{\sigma\in S_{N-n}}1\nonumber\\
&& = (1-\epsilon)^N \sum_{n=0}^N \frac{1}{n!}\left(\frac{\epsilon}{1-\epsilon}\right)^n \ge  (1-\epsilon)^N
\end{eqnarray}
(where for small $\epsilon$  the lower bound is also numerically very close  to the exact value). Eqs. (\ref{E20}) and (\ref{E23}) tell us  that  if   
$\epsilon\le 1 - (1- \varepsilon)^\frac{1}{N}= (1+|O(\varepsilon)|)\varepsilon/N$,  i.e., for $\varepsilon\ll 1$ if  
\be
\epsilon \le \frac{ \varepsilon}{N},
\en{E24}
is satisfied,  the  Boson Sampling with partially distinguishable bosons   with the distinguishability function of  Eq. (\ref{E19})  is $\varepsilon$-close in the total variation distance to the ideal  Boson Sampling.     Assuming the  equivalence  between the two models of imperfections in their effect on the computational complexity of Boson Sampling, Eq. (\ref{E24}) applies  also to  the noisy Boson Sampling.  Note that Eq. (\ref{E24}) agrees with the conclusion of Ref. \cite{KK}  on noise stability of Boson Sampling  for  noise amplitudes $\epsilon = o(1/N)$. This agreement supports our conjecture on the equivalence of the effect of noise to that of  distinguishability on Boson Sampling. However, in section \ref{sec7}  we will show that condition (\ref{E24}) is not \textit{necessary} for classical hardness of noisy Boson Sampling.

 \section{ Efficient classical simulation of the noisy Boson Sampling with finite noise  }
\label{sec6}

In this section we further explore   the connection of the  noisy Boson Sampling to  that with partially distinguishable bosons, discussed in section \ref{sec5}. This connection allows us to  prove theorem 1 of section \ref{secOB}.   It should be stressed, that our proof does not depend on the conjectured equivalence of the effect of two models of imperfection on  Boson Sampling. The main  utility of the above connection  is to simplify some    technical calculations, that  are much easier for the partial distinguishability model than for the noise model, but,  as we show, the results apply to the latter model as well. Most of the  technical calculations,  employed below for the Boson Sampling model with partially distinguishable bosons, have been   performed before in Ref. \cite{VS14}, in  the no-collision regime.  Below we consider  this regime only and heavily rely on the results  of Appendix B of Ref. \cite{VS14}. Moreover, at the end of this section, we provide an additional argument supporting    the conjectured equivalence of the noisy Boson Sampling model Eq. (\ref{E9}) and that with partially distinguishable bosons  Eq. (\ref{E11}) in terms of the classical simulatability. 

Let us first derive  an equivalent, much simpler,  representation of the distinguishability function of Eq. (\ref{E19}). We will use the indicator function $I_{D_{k_{1},\ldots,k_n} }(\sigma)$ of  derangements of $k_1,\ldots,k_n$, i.e., the class of permutations of this set not having  fixed points  (e.g., Ref.  \cite{Stanley}), and the identities: 
\begin{eqnarray}
\label{REL}
&& \sum_{k_1\ldots k_n} I_{ D_{k_{1},\ldots,k_n} }(\sigma)  = \delta_{C_1(\sigma),N-n},\\
&& \prod_{\alpha=1}^s \delta_{k_{\sigma(\alpha)},k_\alpha}  = \sum_{m=s}^N \delta_{C_1(\sigma),m}\sum_{k_{s+1}\ldots k_m} I_{ D_{k_{m+1},\ldots,k_N} }(\sigma), \nonumber
\end{eqnarray}
where the summation in the first line  is over all  $s$-dimensional subsets $k_1,\ldots,k_n $ of $1,\ldots,N$,   $C_1(\sigma)$ is  the total number of fixed points of permutation $\sigma$,   in the second line $k_{s+1},\ldots,k_m$ is the subset of $1,\ldots,N$ consisting  of  the fixed points of $\sigma$ additional to  $k_1,\ldots,k_s$ on the l.h.s..

An equivalent representation of $J(\sigma)$ of Eq. (\ref{E19}) is obtained by performing summation with use  the identities of Eq. (\ref{REL}),  we get 
\begin{eqnarray}
\label{E25} 
&& J(\sigma) = \sum_{s=0}^N \epsilon^s(1-\epsilon)^{N-s}  \sum_{k_1\ldots k_s} \prod_{\alpha=1}^s \delta_{k_{\sigma(\alpha)},k_\alpha} 
\nonumber\\
&& = \sum_{s=0}^N \epsilon^s(1-\epsilon)^{N-s} \sum_{k_1\ldots k_s} \sum_{m=s}^N\sum_{k_{s+1}\ldots k_m} I_{ D_{k_{m+1},\ldots,k_N} }(\sigma)\nonumber\\
&& = \sum_{m=0}^N \sum_{k_1\ldots k_m}\sum_{s=0}^m \binom{m}{s}\epsilon^s(1-\epsilon)^{N-s} I_{ D_{k_{m+1},\ldots,k_N} }(\sigma)\nonumber\\
&&=  \sum_{m=0}^N \sum_{k_1\ldots k_m} (1-\epsilon)^{N-m}I_{ D_{k_{m+1},\ldots,k_N} }(\sigma)\nonumber\\
&& = (1-\epsilon)^{N-C_1(\sigma)},\nonumber\\
&&
\end{eqnarray}
where in the intermediate steps  we  have   introduced the total number ($m$) of fixed  points of  permutation $\sigma$ and  the fixed points themselves $k_1,\ldots, k_m$.

Eq. (\ref{E25}) means that  in the model of Eq. (\ref{E17})  permutation $\sigma$  is  weighted   according to  the number  of derangements  $N-C_1(\sigma)$, thus a permutation with more fixed points $C_1(\sigma)$ contributes with a  larger  weight to  the output probability.

There is another   input state with  partially distinguishable  bosons, different from that of Eq. (\ref{E17}),  giving the same distinguishability function (\ref{E25}). This input state consists of bosons  in pure internal states $|\psi_k\rangle$ with a uniform overlap $\langle \psi_k|\psi_l\rangle = 1-\epsilon$ for $k\ne l$.  Precisely this model was used previously  \cite{R1,RSP} in the proof of classical simulatability of  Boson Sampling with partially distinguishable bosons.  This simplifies our task, as we can follow the previous approach.  The main  step is to consider the following auxiliary version of an imperfect Boson Sampling with partially distinguishable bosons, where   many-boson quantum  interference is allowed only to some  order $R$  
\begin{eqnarray}
\label{E26}
&& \tilde{J}(\sigma) = (1-\epsilon)^{N-C_1(\sigma)}\theta(C_1(\sigma)-N+R),\nonumber\\
&& \theta(m) = \left\{\begin{array}{cc}1, & m\ge 0, \\ 
0, & m < 0.\end{array} \right.
\end{eqnarray}
Now we consider the output probability distributions corresponding to the above two models of partial distinguishability, $J$ of  Eq. (\ref{E25}) and $\tilde{J}$ of Eq. (\ref{E26}). Boson Sampling with   $J$ of Eq. (\ref{E25}) in the no-collision regime has the output probability distribution  given by   Eq. (\ref{E11}),    which we reproduce here  (omitting the input ports $1,\ldots,N$, for simplicity)
\begin{eqnarray}
\label{EQ1}
 &&p(l_1\ldots l_N) = \sum_{n=0}^N B_n(1-\epsilon) \binom{N}{n}^{-1}   \\
 && \times \sum_{{j_1}\ldots {j_n}} \sum_{k_1\ldots k_n}   p^{(U)}_q(l_{j_1}\ldots l_{j_n}|k_1\ldots k_n) \nonumber\\
 &&\times p^{(U)}_{cl}(l_{j_{n+1}}\ldots l_{j_N}|k_{n+1}\ldots k_N) .\nonumber
 \end{eqnarray}
To obtain the output probability distribution for  $\tilde{J}$ of Eq. (\ref{E26}),  we rewrite the $\tilde{J}$-function as the second line of  Eq. (\ref{E19}) with the right hand side multiplied by $\theta(C_1(\sigma)-N+R)$, i.e., 
\be
\widetilde{J}(\sigma)\! = \!\sum_{n=0}^N (1\!-\!\epsilon)^n\epsilon^{N-n}  \!\sum_{k_1\ldots k_n}\sum_{\tau\in S_{n}} \theta(C_1(\tau)-n+R)\delta_{\sigma,\tau\otimes I},
\en{EQ2}
where we have used that $C_1(\sigma) = N-n +C_1(\tau)$ for $\sigma = \tau\otimes I$. 
Comparing    with Eq. (\ref{E19}) we obtain   from Eq. (\ref{EQ1}) the probability distribution for $\tilde{J}$ of Eqs. (\ref{E26}) and (\ref{EQ2}), which we will call the auxiliary Boson Sampling  model with partially distinguishable bosons,
\begin{eqnarray}
\label{EQ3}
 &&\tilde{p}(l_1\ldots l_N) = \sum_{n=0}^N B_n(1-\epsilon)  \binom{N}{n}^{-1} \\
 && \times \sum_{{j_1}\ldots {j_n}} \sum_{k_1\ldots k_n}   \tilde{p}^{(U)}_q(l_{j_1}\ldots l_{j_n}|k_1\ldots k_n),\nonumber\\
 &&\times p^{(U)}_{cl}(l_{j_{n+1}}\ldots l_{j_N}|k_{n+1}\ldots k_N) \nonumber
 \end{eqnarray}
where now 
\begin{eqnarray}
\label{EQ4}
&&\tilde{p}^{(U)}(l_{j_1}\ldots l_{j_n}|k_1\ldots k_n) \\
&& =\sum_{\sigma_1,\sigma_2}\theta(C_1(\sigma_1\sigma^{-1}_2)-n+R) \prod_{\alpha=1}^n U^*_{k_{\sigma_1(\alpha)},l_{j_\alpha}}U_{k_{\sigma_2(\alpha)},l_{j_\alpha}},\nonumber
\end{eqnarray}
with the double summation over $\sigma_{1,2}\in S_n$. The   $\theta$-function in Eq. (\ref{EQ4})  cuts-off   the many-boson interferences  at  the order $R$,  since there  must be  at least $n-R$ fixed points of $\tau$ (see for more details  Ref. \cite{Ninter}). 

Note that  by the results of  sections \ref{sec3} and \ref{sec5}, if in Eqs. (\ref{EQ1}) and (\ref{EQ3}) we replace the classical probability factors by their  averages in a  random network $U$, i.e., $\langle  p^{(U)}_{cl}(l_{j_{n+1}}\ldots l_{j_N}|k_{n+1}\ldots k_N)\rangle = \frac{(N-n)!}{M^{N-n}}$ (in the no-collision regime), we obtain the noisy Boson Sampling model of Eq. (\ref{E5}) and, what we will call below, the auxiliary  noisy Boson Sampling model.  

It is known that, on  average in a  random network $U$, the output probability distributions of Eqs. (\ref{EQ1}) and (\ref{EQ3})    can be made arbitrarily  close   in the total variation distance for any finite  distinguishability $\xi = 1-\epsilon$ by selecting such a $R=O(1)$ and that the auxiliary Boson Sampling model of Eq. (\ref{EQ3}) can be  efficiently simulated classically  for $R=O(1)$ \cite{R1}.   In our proof of efficient classical simulatability  of the   noisy Boson Sampling model, however, we will need some  technical details of  derivation of an  upper  bound on the  total variation distance  between two distributions of Boson Sampling with different partial distinguishability   functions \cite{VS14}.  These details allow us to easily extend the proof  of efficient classical simulatability to the noisy Boson Sampling model as well. 

Consider the variation distance $\mathcal{D}$ between the output distributions of Eqs. (\ref{EQ1}) and (\ref{EQ3}) over the no-collision outputs $l_1< \ldots<l_N$ (the outputs with collisions  can be neglected, since   their contribution is vanishing as $O(N^2/M)$ \cite{AA,Bbirthday}). If we average $\mathcal{D}$ over  the  random unitary  $U$, where we can use the  approximation of  its elements  by  i.i.d. Gaussians \cite{AA}, we obtain 
\begin{eqnarray}
\label{E27} 
&&\langle \mathcal{D}\rangle = \frac12 \sum_{l_1\ldots l_N} \langle |p(l_1\ldots l_N ) - \tilde{p}(l_1\ldots l_N )|\rangle\nonumber\\
&& \le \frac12 \sum_{l_1\ldots l_N} \sqrt{\langle \left(p(l_1\ldots l_N ) - \tilde{p}(l_1\ldots l_N)\right)^2\rangle}\nonumber\\
&& =\frac12 \sum_{l_1\ldots l_N} \frac{N!}{M^N} \left[\frac{1}{N!}\sum_{\sigma\in S_N } \left[ J(\sigma) - \tilde{J}(\sigma)\right]^2\chi(C_1(\sigma))\right]^\frac12\nonumber\\
&& = \frac{1}{2}\left[\frac{1}{N!}\sum_{\sigma\in S_N } \left[ J(\sigma) - \tilde{J}(\sigma)\right]^2\chi(C_1(\sigma))\right]^\frac12,
\end{eqnarray} 
where  we have used that for any real random variable $X$   $\langle |X|\rangle \le \sqrt{\langle X^2\rangle}$~\footnote{Take two mutually independent copies of $X$, say $X_1$ and $X_2$, and expand $\langle (|X_1|- |X_2|)^2\rangle \ge 0$.},  that the average probability is equal to the inverse of  the number of no-collision output configurations (approximated in the no-collision regime by  $ {M^N}/{N!}$) and that  the average of the squared difference  reads 
\begin{eqnarray}
\label{E28}
&& \langle \left(p(l_1\ldots l_N ) - \tilde{p}(l_1\ldots l_N)\right)^2\rangle \nonumber\\
&& = \frac{N!}{M^{2N}}\sum_{\sigma\in S_N} \left[ J(\sigma) - \tilde{J}(\sigma)\right]^2\chi(C_1(\sigma)),
\end{eqnarray} 
with  
\be
\chi(n) \equiv \sum_{\tau\in S_n} 2^{C_1(\tau)}= n!\sum_{k=0}^n \frac{1}{k!}. 
\en{E29}
The derivation of Eq. (\ref{E28})  just repeats that of   Appendix B in   Ref. \cite{VS14}, where one only has to replace the distinguishability function of the ideal Boson Sampling $\tilde{J}(\sigma)=1$ by that of the  model (\ref{E26}).

The right hand side of Eq. (\ref{E28}) can be easily estimated. Introducing the variable  $s \equiv C_1(\sigma)$ and   using the expressions for $J$ and $\tilde{J}$,  Eqs.  (\ref{E25}) and (\ref{E26}),  we obtain 
\begin{eqnarray}
\label{E30}
 &&\frac{1}{N!}\sum_{\sigma\in S_N} \left[ J(\sigma) - \tilde{J}(\sigma)\right]^2\chi(C_1(\sigma))
\nonumber\\
&&= \frac{1}{N!} \sum_{s=0}^{N-R-1}\chi(s) (1-\epsilon)^{2[N-s]}\sum_{\sigma\in S_N}\delta_{C_1(\sigma),s} 
\nonumber\\
&&< \sum_{s=0}^{N-R-1}(1-\epsilon)^{2[N-s]}\left( 1 + \frac{e}{(N-s+1)!}\right)
\nonumber\\
&&< \left( 1 + \frac{e}{(R+2)!}\right) \sum_{s=0}^{N-R-1}(1-\epsilon)^{2[N-s]} 
\nonumber\\
&&< \left( 1 + \frac{e}{(R+2)!}\right)  \frac{(1-\epsilon)^{2[R+1]}}{1-(1-\epsilon)^2},
 \end{eqnarray}
where  we use  $\chi(s) <  s! e$ and a bound on  the  total  number of derangements with  $N-s$ elements \cite{Stanley} 
\[
 \sum_{\sigma\in S_N}\delta_{C_1(\sigma),s}= \frac{N!}{s!}\sum_{i=0}^{N-s}\frac{(-1)^i}{i!}, 
 \]
following from  
\[
 \sum_{i=0}^{N-s}\frac{(-1)^i}{i!} < \left\{\begin{array}{cc} e^{-1}, & N-s = odd, \\ e^{-1} + \frac{1}{(N-s+1)!}, & N-s = even\end{array} \right.
 \] 
 (the sum of two consecutive terms is always positive in the odd case, the even case  is reduced to the odd one by adding  the term  with $i=N-s+1$).
Substitution of the expression in Eq. (\ref{E30}) into Eq. (\ref{E27}) gives an upper bound for the average total  variation distance in the no-collision regime  (up to a vanishing term $O(N^2/M)$) between the distributions in Eqs. (\ref{EQ1}) and (\ref{EQ3})  
\be
\langle \mathcal{D}\rangle < \frac12 \left(1 + \frac{e}{(R+2)!}\right)^\frac12\frac{(1-\epsilon)^{R+1}}{\sqrt{\epsilon(2-\epsilon)}}. 
\en{E31} 
We obtain that for any  $\varepsilon>0$  and the cut-off order of the many-boson interference  $R$  given by  
\be
R = \frac{\ln\left(\frac{\sqrt{2}}{\varepsilon \sqrt{\epsilon}}\right)}{\ln\left(\frac{1}{1-\epsilon}\right)}
\en{E32}
 the average variation distance  (\ref{E31}) between the probability distributions of Eqs. (\ref{EQ1}) and (\ref{EQ3})  is bounded  as  $\langle \mathcal{D}\rangle < \varepsilon/2$. It is easy to see that for any given error $\varepsilon$,  to have $R=O(1)$, as $N\to \infty$, requires that $\epsilon$ is bounded from below away from zero, i.e., $\epsilon =  \Omega(1)$.

The remaining step is to  show that the upper bound   given in   Eq. (\ref{E31})    on the total variation distance between the output distributions of  Boson Sampling model  with partially distinguishable bosons,  Eq. (\ref{EQ1}),  and    the auxiliary   model admitting efficient classical simulations,   (\ref{EQ3}),  remains valid if we  make in  these two  models the same transformation:  replace the classical probability factors in the respective output distributions  by the  average value in a random network $U$. As is noted above, on such a replacement the model of  Eq. (\ref{EQ1})  becomes the noisy Boson Sampling model of Eq. (\ref{E5}), whereas the other  model obviously  continues to allow for  efficient   classical simulations (in this case,   there is  no need    to simulate the classical probabilities).      To the above  goal, we need some facts established previously \cite{VS14} on the main ingredient of the upper bound -- the variance of the difference in respective probabilities. The variance is  given by  Eq. (\ref{E28}), since in the no-collision regime the average probability is uniform over the output ports,   independently  of the distinguishability function  \cite{VS14}.   Replacing the classical probability factor    by its average in the output distributions  \textit{can only decrease}  the variance of the difference, since the   classical  factors   in different terms in the respective output probability distributions, i.e., in the sums  in Eqs.  (\ref{EQ1}) and (\ref{EQ3}), contain  independent    random variables,  due to mutual independence of  elements of  a random unitary matrix in the no-collision regime \cite{AA}.  Therefore,  we have proven that Eq. (\ref{E31}) gives the required bound  on the total variation distance between noisy Boson Sampling model of Eqs. (\ref{E5}) or (\ref{E1}) and  an  auxiliary model   that  can be efficiently simulated classically.  

 The  classical simulatability  result  can be formulated as follows \cite{R1}:  the number of classical computations required to simulate the  noisy Boson Sampling model Eq. (\ref{EQ1})   to a given error $\varepsilon$ and with the probability of success at least $1-\delta$ for any $\delta >0$  is  a polynomial function of  $(N,1/\varepsilon,1/\delta)$ for    $\epsilon =  \Omega(1)$.    Hence, we have proven   theorem 1 of section \ref{Outline}.
An explicit  algorithm for  classical simulations of  Boson Sampling with partially distinguishable bosons  of Eq. (\ref{E26}) is the same as in   Refs. \cite{R1,RSP}  (see the comment in the paragraph above Eq. (\ref{E26})).  

Now let us examine  by how much  the variance of the difference  in Eq. (\ref{E28})  decreases after replacing the classical probability factors in Eqs. (\ref{EQ1}) and (\ref{EQ3}) by the  average values.   The change of the variance in the Gaussian approximation comes from the  fourth-order  moments $\langle |U_{kl}|^4\rangle= 2/M^2$ (appearing when  averaging a squared classical probability factor)  being  replaced by $(\langle |U_{kl}|^2\rangle)^2 = 1/M^2$.  But the fourth-order moments that are being replaced are weighted at least by  $\epsilon^2$, since they come from the   noise contribution to output distribution of noisy  Boson Sampling, where each term is    weighted at least by $\epsilon$ (the variance is   quadratic function   in the probability and different matrix elements are mutually independent random variables in the no-collision regime).  The  decrease  in  the relative variance is therefore on the order  $O(\epsilon^2)$. This  very small difference in the variance for  small amplitude of imperfection,  when switching between the noisy Boson Sampling model and that with partially distinguishable bosons,    supports the  conjecture  of section \ref{sec5} on their  equivalence   in terms of    classical hardness.

 \section{How many    ``noise clicks"  in  noisy Boson Sampling?}
\label{sec7}

Let us return to the  output probability distribution of Eq. (\ref{E9}), i.e., the equivalent  representation of the noisy Boson Sampling model.  It replaces the continuous model of  noise by an equivalent one with discrete ``noise clicks" (i.e., lost   bosons compensated by random detector counts). Consider now  the noisy Boson Sampling model with $N$ input bosons,  but with the number of ``noise clicks"  bounded by   $R-1$.  Given   $\varepsilon\ll 1$,  what number of clicks   $R = R(N,\varepsilon)$ is sufficient to approximate the output probability distribution of a given  noisy Boson Sampling  (\ref{E9}) to the  total variation distance $\mathcal{D}\le \varepsilon$?  The  question is answered below and the answer allows to prove theorem 2 of section \ref{Outline} by using the results of Ref. \cite{Brod}.

 The probability distribution of the noisy Boson Sampling model   with the total number of ``noise clicks"  bounded by   $R-1$ is obtained by imposing  a  cut-off on the summation index $ N-R+1\le n\le N$ in  Eq. (\ref{E9}) and rescaling (here we  use $\langle \ldots \rangle$ to denote averaging over the noise $Z$, as  in sections \ref{sec3} and \ref{sec4})
\begin{eqnarray}
\label{E33}
&&\langle p^{(\U)}_R(\m |1\!\ldots\! N)\rangle =  C_R\sum_{n=N-R+1}^{N} B_n(1-\epsilon)  \nonumber\\
&& \times  \binom{N}{n}^{-1} \sum_{k_1\ldots k_n}   \sum_{{\s \subset\m \atop |\s|=n}}  q(\br) p^{(U)}_q(\s |k_1\!\ldots\! k_n),\nonumber\\
\end{eqnarray}
where $C_R = \sum B_n(1-\epsilon)$ for  $ N-R+1\le n\le N$.
 We get the following bound on the  total variation distance between models of Eq. (\ref{E9}) and (\ref{E33})  
\begin{eqnarray}
\label{E34}
&& \mathcal{D} = \frac12 \sum_{\m} |\langle p^{(\U)}(\m |1\!\ldots\! N)\rangle  -\langle p^{(\U)}_R(\m |1\!\ldots\! N)\rangle | \nonumber\\
&& \le  \frac12 \left( \left[\frac{1}{C_R}-1\right]\sum_{n=N-R+1}^{N} B_n(1-\epsilon)  + \sum_{n=0}^{N-R} B_n(1-\epsilon)  \right)\nonumber\\
&& =  \sum_{n=0}^{N-R} B_n(1-\epsilon) =   \sum_{s=R}^{N} B_s(\epsilon),
\end{eqnarray}
  where to perform summation over $\m$  we have used that  the  terms  in Eq. (\ref{E33}) (as well as  in Eq. (\ref{E9}))  are positive and represent  certain probabilities. 
 
One immediate consequence of Eq. (\ref{E34}) is that for noise amplitude $\epsilon = O(1/N)$  and any given small error $\varepsilon$ taking   $R = O(1)$ is sufficient to bound the total variation distance as required, $ \mathcal{D}\le \varepsilon$. Recall that for $\epsilon =O(1/N) $ the binomial distribution $B(\epsilon)$  is approximated by the Poisson distribution with the expected  number of clicks $\langle R\rangle  = \epsilon N =O(1)$, thus bounding the number of clicks by $R$, such that  $R > \langle R\rangle $,  is sufficient to approximate the distribution (i.e., make  the tail in Eq. (\ref{E34}) arbitrarily small). Below this is  formally proven  with the use of Hoeffding's bound    \cite{Hoeffding}. 

Hoeffding's bound (see theorem 1 in Ref. \cite{Hoeffding}), applied here for  the partial sum of the Binomial distribution represented as $N$ i.i.d. trials $x_i\in \{0,1\}$, where $x_i=1$ with probability $\epsilon$,  states that for $\epsilon  < R/N < 1$
 \begin{eqnarray}
\label{E35}
&& \sum_{s=R}^{N} B_s(\epsilon) = \mathrm{Prob}\left(\sum_{i=1}^N x_i\ge R\right)\nonumber\\
&& \le \left(\frac{N\epsilon}{R}\right)^R \left( \frac{1-\epsilon}{1-R/N}\right)^{N-R}.\nonumber\\
  \end{eqnarray}
We conclude that    if  there is such $R$ satisfying   
\be
 \left(\frac{N\epsilon}{R}\right)^R \left( \frac{1-\epsilon}{1-R/N}\right)^{N-R} \le \varepsilon,\quad R >\epsilon N, 
\en{E37} 
then  $\mathcal{D}\le \varepsilon$ as required. Consider the second factor in the first inequality in Eq. (\ref{E37}),  setting  $X\equiv (R/N-\epsilon)/(1-\epsilon)$ we obtain   
\begin{eqnarray*}
&& \left( \frac{1-\epsilon}{1-R/N}\right)^{N-R} = \exp\{-(N-R)\ln(1- X) \} \\
&& \le    \exp\left\{ \frac{(N- R)X}{1-X}\right\} =  \exp\left\{ \frac{(N- R)(R/N-\epsilon)}{1-R/N}\right\} \\
&& =\exp\{R - N\epsilon\},
\end{eqnarray*}
where we have used that  $- \ln(1-X) \le   X/(1-X)$ (here $0< X<1$, under  the second condition in Eq. (\ref{E37})). 
Then,  Eq. (\ref{E37}) follows from the following simpler one
\be
  \left( \frac{R}{e\epsilon N} \right)^R \ge \frac{ e^{-\epsilon N} }{\varepsilon}, \quad R > \epsilon N,
\en{E38}
which  can be satisfied for $\epsilon=O(1/N)$ by choosing  $R = O(1)$.

The existence of  a  finite bound $R=O(1)$  on the total number of ``noise clicks" resulting in arbitrary closeness of the reduced model  Eq. (\ref{E33})   in the total variation distance  to the noisy Boson Sampling model Eq. (\ref{E9}),  and  the  equivalence of the noisy Boson Sampling model (with  bounded ``noise clicks")  to that of the shuffled bosons  model of imperfect Boson Sampling (boson losses compensated by detector dark counts) with a finite number of losses and dark counts  allows us to use  the  results of Ref. \cite{Brod} on the classical  hardness of  such a shuffled bosons model (in the no-collision regime).  We get   that  in the no-collision regime the   noisy  Boson Sampling with the noise amplitude $\epsilon = O(1/N)$ is as hard as the  ideal Boson Sampling for classical simulations. Hence, we have proven theorem 2 of section \ref{Outline}.

In Ref. \cite{KK} it was suggested that for $O(1/N)$-noise the  noisy Boson Sampling model of Eq. (\ref{E1})  must be at a finite total variation distance from the ideal one.  The same is  implied  also by  Eqs. (\ref{E20})-(\ref{E24}) of section  \ref{sec5} under the conjectured equivalence of the noisy Boson Sampling Eq. (\ref{E9}) and that with partially distinguishable bosons Eq. (\ref{E11}),  if the bound in Eq. (\ref{E20})  is  tight for the closeness to the ideal Boson Sampling,  as suggested in Ref. \cite{VS15}. Nevertheless, a finite total variation distance from the ideal Boson Sampling for  such noise   does not   allow, by theorem 2,   an efficient classical simulation of the respective   noisy Boson Sampling model. 

Finally, from section \ref{sec6} and Ref. \cite{KK},  we know  that  for finite  noise $\epsilon =  \Omega(1)$ the noisy Boson Sampling of Eq. (\ref{E1}) can be  efficiently simulated classically.  Consider now   the intermediate  noise amplitude   $\epsilon=o(1)$.  It was  shown that the correlation between the noisy and noiseless outcomes of Boson Sampling  tends to zero   for $\epsilon = \omega(1/N)$ \cite{KK}.  In this case    by Eq. (\ref{E37}) $R$  is unbounded  and the above proof fails.    However,    when also    $\epsilon = o(1)$ (i.e., vanishing noise amplitude),  we can satisfy  Eq. (\ref{E38}), as $N\to \infty$, e.g.,   with   $R =  2\epsilon N$.  Thus in this case    the relative effective number of ``noise clicks" $R/N$ (the relative effect of noise on $N$-boson quantum interference) is  vanishing,  similar as in the classically hard  case  $\epsilon= O(1/N)$.  In view of this fact we conjecture that no efficient    classical  simulation  is possible for  any  small noise amplitude  $\epsilon =o(1)$.

\medskip
\section{Conclusion}
\label{Concl}

 We have found two physically  realistic models of imperfections, the partial distinguishability of bosons and  the boson losses compensated by dark counts of detectors, which lead to the  Gaussian noise model  in Boson Sampling introduced and studied previously in Ref. \cite{KK}. This enables us to precisely  compare previous results on sufficient and necessary conditions for classical hardness of noisy/imperfect realisations of Boson Sampling and present a unified  picture of the effect of noise.    We  find that noisy Boson Sampling with a finite amount of  noise, i.e., with the noise amplitude  $\epsilon =\Omega(1)$ as $N\to \infty$ allows for efficient  classical simulations, which confirms the previous  result  of Ref. \cite{KK}. We also prove that noisy Boson Sampling with noise amplitude  scaling inversely with the total number of bosons $\epsilon = O(1/N)$  retains classical hardness, though, by Ref. \cite{KK},   confirmed by our results, its output distribution is far from the ideal Boson Sampling, as it  is at a constant  total variation distance from the latter.     This  means that closeness to the ideal Boson Sampling  is not the only possible cause of  classical hardness of an imperfect (i.e., experimental) realisation, meaning that  some of the necessary conditions known in the literature, derived under the requirement of a small such total variation distance,  are not actually necessary for classical hardness of a noisy/imperfect  physical realisation. 
 
 The above  facts have  strong  implications for experimental efforts to demonstrate quantum advantage with Boson Sampling.  First of all, our results   significantly reduce the previous sufficient condition on the average probability of error (due to noise)  in network elements, e.g.,    beam-splitters, such that  an experimental  Boson Sampling with $N$ bosons on a $M$-dimensional noisy network remains  still classically hard to simulate, from previous $o(1/(N^2\log M))$ \cite{Arkhipov}  to $O(1/(N\log M))$, corresponding to the sufficient  bound on  noise amplitude $\epsilon = O(1/N)$. However,  our  results  also reveal that  the latter may   not be  necessary  for  classical hardness.    By analysis of the physical effect  of   noise  on many-body quantum interference in Boson Sampling, we are driven to the conjecture that any small noise, i.e.,  with vanishing amplitude $\epsilon = o(1)$ is sufficient for classical hardness of the output distribution, since the ratio of the  effective total number of ``noise clicks" (lost boson compensated by dark counts of detectors)  to the number of bosons $N$ vanishes   as $N\to\infty$. This conjecture implies  that the efficient  classical simulatability threshold lies at   $\epsilon =\Omega(1)$ (finite noise).  If true, this fact would further reduce    the sufficient condition on the  average  probability of error in network elements from the proven here $O(1/(N\log M))$ to just $o(1/\log M)$, i.e., the   error in an elementary block of network should decrease faster than  the inverse of network depth. 
 
Moreover,   an  extension of the Gaussian noise model beyond the usual  no-collision regime has been proposed in  the present work. At least some of our conclusions, such as the analysis leading to the conjecture above,  are valid for general  $M\ge N$,  beyond the no-collision regime.     However,  due to technical reasons (we use previous results with restricted validity)   the proven   results  on the computational  complexity   apply to  the no-collision  regime only.   Efficient classical algorithms due to imperfections  in experimental setup exist also for  Boson Sampling in  arbitrary  regime $M\ge N$   \cite{K1,OB,PRS}. All of these results do not contradict the conjecture that a finite noise amplitude is the threshold of classical simulations of noisy  no-collision Boson Sampling. However, this does not imply that this threshold is the same for no-collision and beyond the no-collision regimes of Boson Sampling.     New   results in this direction indicate   strong dependence of sensitivity of Boson Sampling to noise  on  the  ratio between the total number of bosons and   network size \cite{NEW}. 

 \medskip
\section{Acknowledgements}  
The author   was supported by the National Council for Scientific and Technological Development (CNPq) of Brazil,  grant  304129/2015-1 and by the S\~ao Paulo Research Foundation (FAPESP), grant 2018/24664-9.


\end{document}